\documentclass[]{jfm}

\usepackage{graphicx}
\usepackage{floatrow}
\usepackage{subfig}
\usepackage[justification=justified]{caption}
\usepackage{newtxtext}
\usepackage{newtxmath}
\usepackage{natbib}
\usepackage[framed,numbered,autolinebreaks]{mcode}
\usepackage{hyperref}
\hypersetup{
    colorlinks = true,
    urlcolor   = blue,
    citecolor  = black,
}
\usepackage{bm}
\usepackage{upgreek}

\newcommand{\RomanNumeralCaps}[1]

\title[Drag force of a compressible flow past spheres]{
Drag force of a compressible flow past a random array of spheres
}

\author[M. Khalloufi and J. Capecelatro]{Mehdi Khalloufi\aff{1}
  \corresp{\email{khalloufi.mehdi@gmail.com}
  }
 \and~Jesse Capecelatro\aff{1,2}}

\affiliation{\aff{1}Department of Mechanical Engineering, University of Michigan, Ann Arbor, MI 48109, USA
\aff{2}Department of Aerospace Engineering, University of Michigan, Ann Arbor, MI 48109, USA
}

\begin{document}
\maketitle

\begin{abstract}
We perform particle-resolved simulations of subsonic and transonic flows past random arrays of spherical particles. The Reynolds number is held at $Re{\approx}300$ to ensure the flow remains in the continuum regime. At low volume fractions, the drag force increases sharply near a critical Mach number due to the formation of shock waves and reaches a maximum value when the bulk flow is supersonic. Neighbour-induced hydrodynamic interactions reduce the critical Mach number at higher volume fractions. An effective Mach number is introduced to capture the increase in compressibility effects on drag. A new drag correlation is proposed valid for subsonic and weakly supersonic flow from dilute to moderately dense suspensions.
\end{abstract}


\section{Introduction}
The importance of fluid resistance on the dynamical evolution of particles has led to long-standing efforts in obtaining accurate estimates for the drag of a sphere.
Book II of Isaac Newton's \textit{Principia} was one of the earliest works on the topic, demonstrating the drag force is proportional to the square of the object's speed through the fluid $U$, its cross-sectional area $A$ and the density of the carrier fluid $\rho$; $F_d{=}\rho A U^2C_D/2$. His early experiments showed the drag coefficient of a sphere to be $C_D{\approx}0.5$. Almost two centuries later, \citet{stokes1851effect} derived an analytic solution for the drag on a sphere moving through an incompressible  fluid in the limit $Re{\equiv}\rho U D/\mu{\ll}1$. The resulting drag coefficient, $C_D{=}24/Re$, acts as the basis for nearly all drag laws that have been proposed thereafter.

A culmination of experimental and semiempirical studies from the 20th century yielded reliable estimates of $C_D$ for a sphere in incompressible flows up to $Re{<}10^5$~\citep[e.g.][]{oseen1910uber,goldstein1929steady,schiller1933fundamental,clift1970motion,abraham1970functional}. \citet{miller1979sphere} compiled available data for the drag on spheres over a wide range of Reynolds and Mach numbers ($M$) using data from 18th and 19th century cannon firings and free-flight measurements from aeroballistic ranges. They showed that drag tends to increase monotonically with $M$ at high values of $Re$. At low $Re$ and moderate to high $M$, non-continuum effects become important and drag  approaches the free-molecular limit. In transonic conditions ($0.9{<}M{<1.1}$), $C_D$ increases sharply due to the emergence of a bow shock in the vicinity of the sphere.

Using available experimental and theoretical data at the time, \citet{henderson1976drag} developed a $Re$- and $M$-dependent drag correlation for subsonic and supersonic flows, and linearly interpolated the data for transonic conditions. 
\citet{loth2008compressibility} later developed a drag coefficient by separating the flow into a rarefaction-dominated regime for $\Rey{\lessapprox}45$ and a compression-dominated regime when $\Rey{\gtrapprox}45$. 
\citet{parmar2010improved} assessed the models of \citet{henderson1976drag} and \citet{loth2008compressibility} using data collected by \citet{bailey1976sphere}. Both models were found to yield significant errors near the transonic regime. They proposed an improved model for $M{\le}1.75$ by decomposing $C_D$ into three correlations for subcritical ($M{<}0.6$), intermediate and supersonic regimes.

With the advent of high-performance computing, direct numerical simulations (DNS) are beginning to shed new light on this topic.  \citet{loth2021supersonic} combined DNS data of \citet{nagata2020direct} with rarefied-gas simulations and an expanded experimental dataset to refine $C_D$. It was shown to be approximately twice as accurate compared to the correlations of \citet{loth2008compressibility} and \citet{parmar2010improved} at moderate Mach numbers, and showed improvement to Loth's original model at $M{>}2$. In the same year, \citet{singh2022general} developed a physics-based expression for $C_D$ that incorporates rarefied effects, low-speed hydrodynamics and shock-wave physics to accurately model drag on a sphere for a wide range of $Re$ and $M$.

\begin{figure}
     \centering
         \includegraphics[width=0.4\columnwidth]{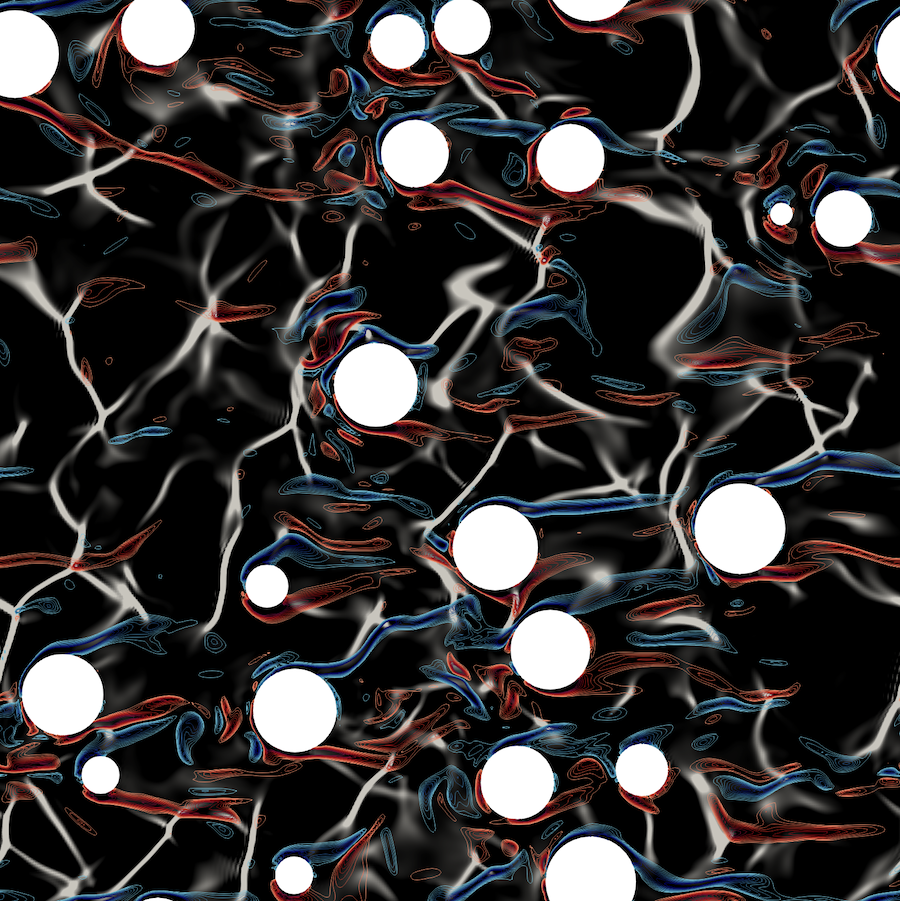}
         \caption{Slice in the $x{-}y$ plane of a homogeneous assembly of particles with $Re=300$, $M=1.0$ and $\phi=0.1$ showing fluid dilatation $-5{<}\nabla\bm{\cdot}\bm{u}{<}0$ (black/white) and contours of the $z$-component of vorticity $-10{<}\omega_z{<}10$ (red/blue).}
         \label{fig:dilatation}
\end{figure}

Yet, all of these efforts have considered flow past an \textit{isolated} sphere.
Over the past two decades, many correlations have been proposed for particle assemblies in incompressible flows ~\cite[e.g.][]{hill2001first,beetstra2007drag,tenneti2011drag,tang2015new}, while much less attention has been paid to compressible flows. At finite Mach numbers and volume fractions ($\phi$) the flow can accelerate to supersonic speeds in the interstitial space between particles, resulting
in large values of fluid dilatation (see Fig.~\ref{fig:dilatation}). Flows with finite values of $Re$, $M$ and $\phi$ are encountered in numerous environmental and engineering applications, such as volcanic eruptions~\citep{lube2020multiphase}, coal dust explosions~\citep{houim2015numerical}, solid propellant combustion~\citep{kulkarni1982review} and jet-induced cratering during spacecraft landing~\citep{mehta2013thruster,balakrishnan2021fluid,capecelatro2022modeling}.
In recent years, DNS of compressible flows past assemblies of particles have started to come online \citep[e.g.][]{regele2014unsteady,mehta2018propagation,theofanous2018shock,osnes2019computational,shallcross2020volume}. However, these studies typically consider shock--particle interactions, which introduce challenges in developing drag correlations due to the lack of statistical stationarity and homogeneity. In what follows, the drag force obtained from particle-resolved simulations of homogeneous compressible flows past random arrays of spherical particles is quantified for $0.1{\le}M{\le}1.2$ and $0.02{\le}\phi{\le}0.4$. A new drag correlation is proposed that introduces the notion of a volume fraction-dependent effective Mach number to capture increased compressibility effects caused by particles.

\section{Simulation methodology}
\subsection{Governing equations and numerical methods}
We consider rigid spherical particles suspended in an unbounded viscous compressible gas. The Navier--Stokes equations governing the gas phase are given by
\begin{equation}\label{eq:density}
\frac{\partial \rho}{\partial t}+\nabla\bm{\cdot}\left(\rho \bm{u} \right)=0,
\end{equation}
\begin{equation}\label{eq:momentum}
\frac{\partial \rho \bm{u}}{\partial t} + \nabla \bm{\cdot} ( \rho \bm{u}\bm{u} + p \bm{I}  - \bm{\tau}) = 0,
\end{equation}
and
\begin{equation}\label{eq:energy}
\frac{\partial \rho E}{\partial t} + \nabla \bm{\cdot} ( \{\rho E+ p\} \bm{u} - \bm{u} \bm{\cdot} \bm{\tau}-\kappa\nabla T) = 0,
\end{equation}
where $\rho$ is the gas-phase density, $\bm{u}{=}(u,v,w)^{\sf T}$ is the velocity vector, $E$ is the total energy, $p$ is pressure, $T$ is temperature, $\kappa$ is the thermal conductivity and $\bm{I}$ is the identity tensor. The fluid is taken to be a caloric perfect gas. All variables are non-dimensionalised with the ambient density $\rho_\infty^\star$, particle diameter $D^\star$, dynamic viscosity $\mu_\infty^\star$, heat capacity at constant pressure $C_p^\star$ and speed of sound $c_\infty^\star{=}\sqrt{\gamma p_\infty^\star/\rho_\infty^\star}$, where $\gamma$ is the ratio of specific heats. The superscript $\star$ and subscript $\infty$ denote a dimensional and reference quantity, respectively.
The viscous stress tensor is given by
$\bm{\tau} {=} \mu( \nabla \bm{u} + \nabla \bm{u}^{\sf T})/\Rey_c+ \lambda(\nabla \bm{\cdot} \bm{u}) \bm{I}/\Rey_c$, where $\mu$ is the dynamic viscosity, $\lambda{=}\beta-2\mu/3$ is the second coefficient of viscosity and $\beta$ is the bulk viscosity. $Re_c {=} Re / M$, where $Re {=} \rho_\infty^\star U_\infty^\star D^\star / \mu_\infty^\star$ is the Reynolds number based on the free-stream velocity $U_\infty^{\star}$ and $M {=} U_\infty^{\star} / c^{\star}_{\infty}$ is the reference Mach number. The viscosity follows a power law $\mu{=}\left[(\gamma-1)T\right]^n$, where $n {=} 0.666$ is a model for air. The thermal conductivity $\kappa{=}\mu/(Re_c  Pr)$ is assumed to have the same temperature dependence as viscosity and the Prandtl number is held constant at $Pr{=}0.7$.

Spatial derivatives are discretized using sixth-order central finite difference operators and time advancement is achieved using a fourth-order Runge--Kutta scheme, resulting in the usual Courant--Friedrichs--Lewy (CFL) restrictions on the simulation time step, $\Delta t$. Kinetic energy preservation is achieved using a skew-symmetric-type splitting of the inviscid fluxes~\citep{pirozzoli2011stabilized}, providing nonlinear stability at low Mach number. High-order energy-stable dissipation operators based on a sixth-order derivative~\citep{mattsson2004stable} are used to damp the highest wavenumber components supported by the grid. 

Localized artificial diffusivity is used as a means of shock capturing following the `LAD-D2-0' formulation in \citet{kawai2010assessment}. The bulk viscosity and thermal conductivity are augmented according to $\beta{=}\beta_f+\beta^*$ and $\kappa{=}\kappa_f+\kappa^*$, where $f$ subscripts and asterisks denote fluid and artificial transport coefficients, respectively. The artificial transport coefficients take the form $\beta^*{=}C_\beta\overline{\rho f_{sw}|\nabla^4\theta|}\Delta^6$ and $\kappa^*{=}C_\kappa\overline{\rho c |\nabla^4 e|/ T}\Delta^5$, where $\theta{=}\nabla\bm{\cdot}\bm{u}$, $e{=}(\gamma-1)^{-1}p/\rho$, $C_\beta{=}1$, $C_\kappa{=}0.01$ and $\Delta$ is the grid spacing. The overbar denotes a truncated 9-point Gaussian filter \citep{cook2004high}. Fourth derivatives are approximated via a sixth-order compact (Pad\`e) finite-difference operator \citep{lele1992compact}. To limit the artificial bulk viscosity to regions of high compression (shocks), we employ the sensor of \citet{hendrickson2018improved}, given by $f_{sw}{=}\min\left(4/3H(-\theta)\times\theta^2/(\theta^2+\Omega^2+\epsilon),1\right)$, where $H$ is the Heaviside function, $\epsilon{=}10^{-32}$ is a small positive constant to prevent division by zero and $\Omega{=}\max\left(|\nabla\times\bm{u}|,0.05 c/\Delta \right)$ is a frequency scale that ensures the sensor goes to zero where vorticity is negligible.

A ghost-point immersed boundary method originally proposed by \citet{mohd1997combined} and later extended to compressible flows by \citet{chaudhuri2011use} is employed to enforce boundary conditions at the surface of each particle on a Cartesian grid. Values of the conserved variables at ghost points residing within the solid are assigned after each Runge--Kutta sub-iteration to enforce adiabatic no-slip boundary conditions. To avoid introducing unphysical discontinuities near the immersed interface, $\beta^*$ and $\kappa^*$ are defined at every grid point within the domain (fluid and solid), but values inside the solid are excluded when computing the CFL.

\begin{figure}
     \centering
         \includegraphics[width=0.5\columnwidth]{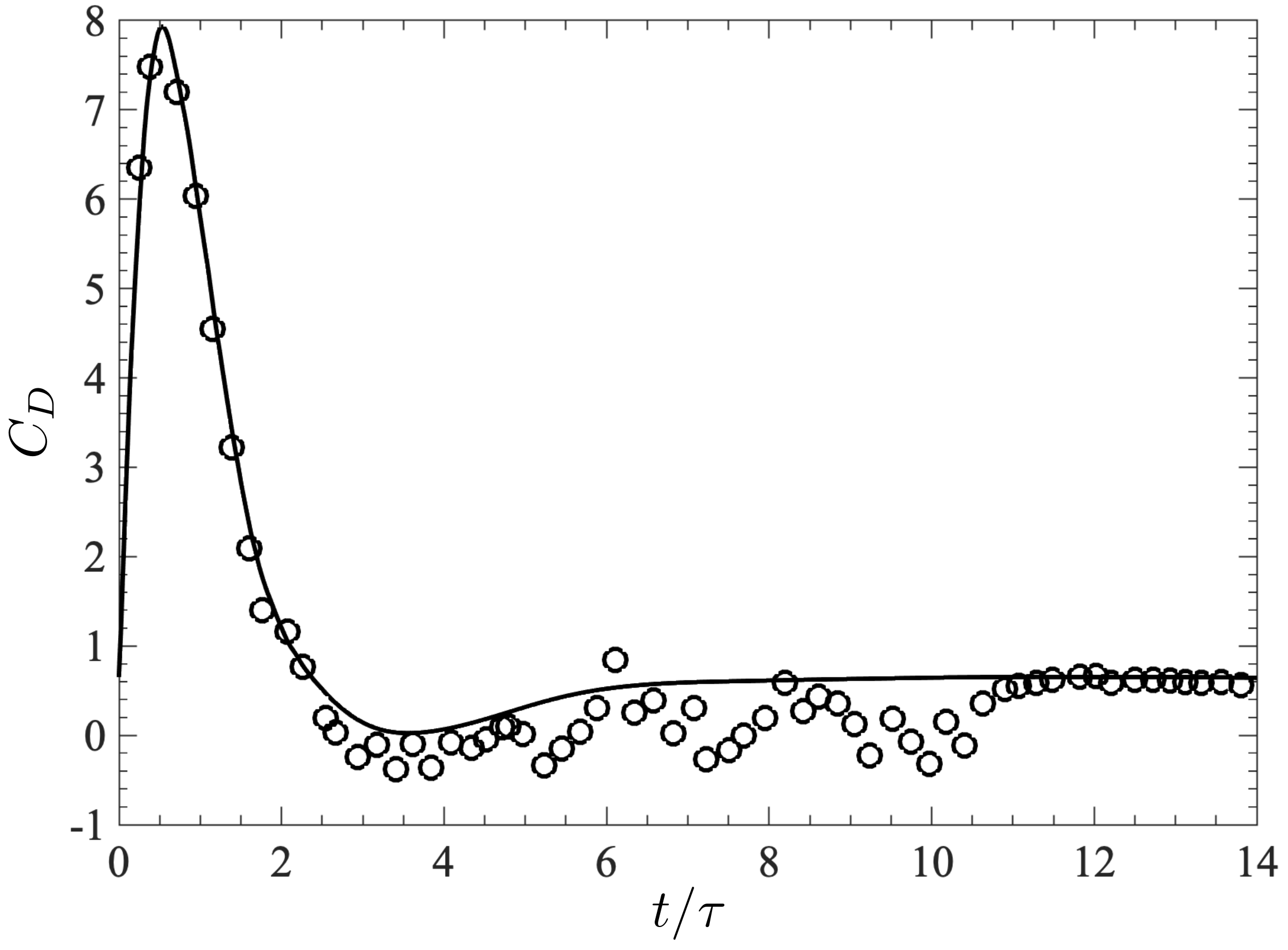}
         \caption{Unsteady drag coefficient of a sphere exerted by a shock wave with Mach number 1.22. Simulation with $\Delta=D/40$ ($-$) and experiment by \citet{sun2005unsteady} ($\circ$).}
         \label{fig:validation}
\end{figure}

The drag force is computed via volume integration of the fluid stresses according to $\bm{F}_d{=}\iiint_{\Omega}\nabla\bm{\cdot}\left(-p\bm{I}+\bm{\tau}\right){\rm d}V{\approx}\sum_{k\in\Omega}\nabla\bm{\cdot}\left(-p\mathcal{I}+\bm{\tau}\right)\Delta$, where the volume is evaluated over all grid points $k$ within the particle of volume $\Omega$. Figure~\ref{fig:validation} shows the unsteady drag coefficient, $C_D{=}F_d/(0.5\rho_\infty U_\infty^2 A)$, exerted by a shock with Mach number 1.22 and $Re{=}4.9\times10^5$, where $F_d=\bm{F}_d\bm{\cdot}\hat{\bm{x}}$. Time is normalized by $\tau{=}D/(2\sqrt{p_\infty/\rho_\infty})$. Good agreement with experimental data of \citet{sun2005unsteady} is observed using $\Delta{=} D/40$. Further details on the simulation configuration used for validation can be found in \citet{shallcross2022explicit}. We note that the rapid change in $C_D$ occurs when the shock wave passes over the particle. This work is primarily concerned with the `quasi-steady' drag coefficient, i.e. the drag when $t\gg\tau$. 

\subsection{Flow configuration}
We consider a homogeneous suspension of stationary particles with diameter $D$ in a three-dimensional periodic box of length $L$ (see Fig.~\ref{fig:dilatation}). The number of particles is chosen based on $N_p{=}6\phi/\pi \left(L/D\right)^3$, where $\phi$ is the solid volume fraction. The mean Reynolds number is set by a constant mass flow rate proportional to $(1-\phi)\rho_\infty U_\infty$ and the ambient pressure is varied to independently enforce the free-stream Mach number, given respectively as
\begin{equation}
    Re_m=(1-\phi)\frac{\rho_\infty U_\infty D}{\mu_\infty}\quad\text{and}\quad  M=\frac{U_\infty}{c_\infty}.
\end{equation}
Here, $U_\infty{=}\langle\rho\mathcal{I} u\rangle/\langle\rho\mathcal{I}\rangle$ is the mean velocity, where angled brackets denote an average in space and time and $\mathcal{I}(\bm{x})$ is the indicator function that is unity if the point $\bm{x}$ lies in the gas phase and zero otherwise, with $\langle\mathcal{I}\rangle{=}1-\phi$. In this work, a phase average is defined as $\overline{(\cdot)}{=}\langle\mathcal{I}(\cdot)\rangle/\langle\mathcal{I}\rangle$ and a Favre average is defined as $\widetilde{(\cdot)}{=}\langle\rho\mathcal{I}(\cdot)\rangle/\langle\rho\mathcal{I}\rangle$. The volume fraction and Mach number are varied between $0.02{\le}\phi{\le}0.4$ and $0.1{\le} M{\le}1.2$. The gas has constant ratio of specific heats $\gamma{=}1.4$ and  consequently the Knudsen number can be expressed as $Kn{=}\sqrt{\pi \gamma/2}M/\Rey$, where $\Rey=\Rey_m/(1-\phi)$. We take $\Rey_m{=}300$ to ensure $Kn{<}.005$ over the range of cases so that the continuum approximation holds.

To enforce a constant mass flow rate, a uniform body force $f_u{=}(\rho_\infty U_\infty-\overline{\rho u})/\Delta t$ is added to the right-hand side of the $x$-component of momentum \eqref{eq:momentum} at each stage of the time integrator. 
Accordingly, the power spent $uf_u$ is added to the right-hand side of \eqref{eq:energy}. Due to periodicity and long time horizons considered, spurious mass errors might accumulate when enforcing momentum \citep{coleman1995numerical}. Therefore, a similar source term is added to \eqref{eq:density} of the form $f_\rho{=}(\rho_\infty-\overline{\rho})/\Delta t$. Finally, thermal forcing is added to the right-hand side of \eqref{eq:energy} to prevent accumulation of heat via $f_T {=}\rho/\gamma(T_\infty-\overline{T})/\Delta t$.

The domain is discretized with uniform grid spacing $\Delta$. The grid spacing and domain size were varied between $20{\le}D/\Delta{\le}90$ and $4{\le}L/D{\le}12.8$ to assess convergence. At low volume fractions, drag and turbulent kinetic energy were found to converge with $L/D{=}10$ and $D/\Delta{=}40$. As the interparticle spacing reduces with increasing volume fraction, addition resolution was deemed necessary. The results presented in the following section are obtained using 40 grid points per diameter for $\phi{<}0.3$ and 60 grid points per diameter for $\phi{\geqslant}0.3$.

\section{Results}\label{sec:results}
\begin{figure}
  \centering
  \sidesubfloat[]{\includegraphics[height=.25\textwidth]{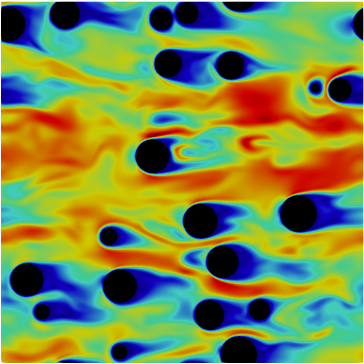}}~
 \sidesubfloat[]{\includegraphics[height=.25\textwidth]{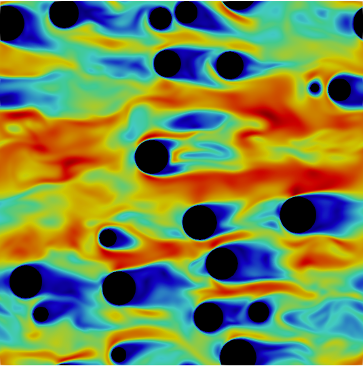}}~
 \sidesubfloat[]{\includegraphics[height=.25\textwidth]{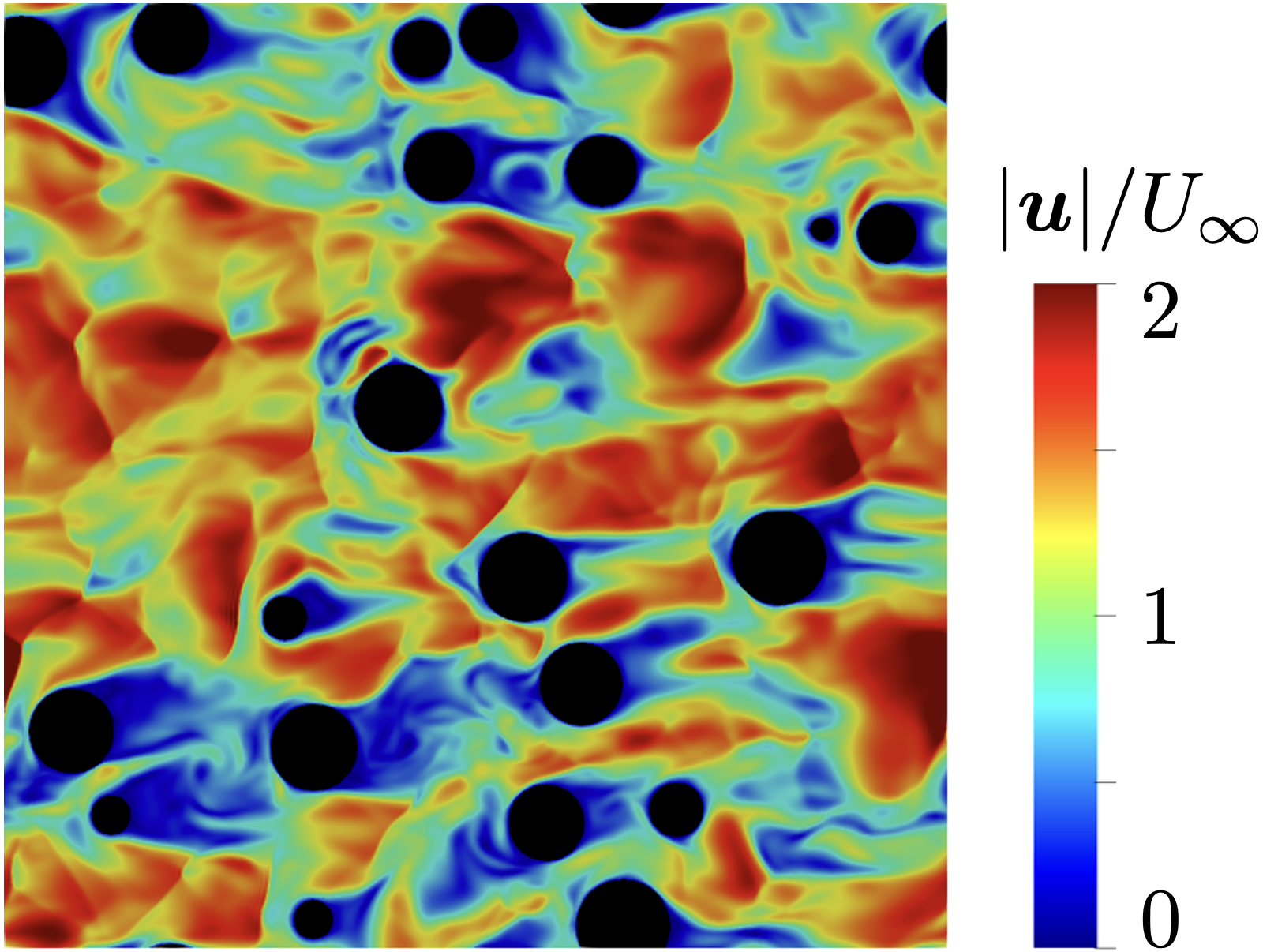}}\\
 \vspace{.05in}
  \sidesubfloat[]{\includegraphics[height=.25\textwidth]{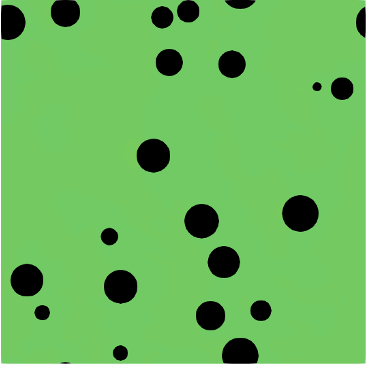}}~
  \sidesubfloat[]{\includegraphics[height=.25\textwidth]{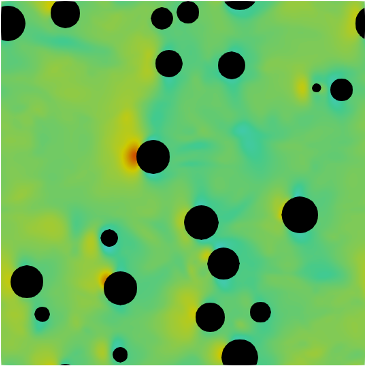}}~
  \sidesubfloat[]{\includegraphics[height=.25\textwidth]{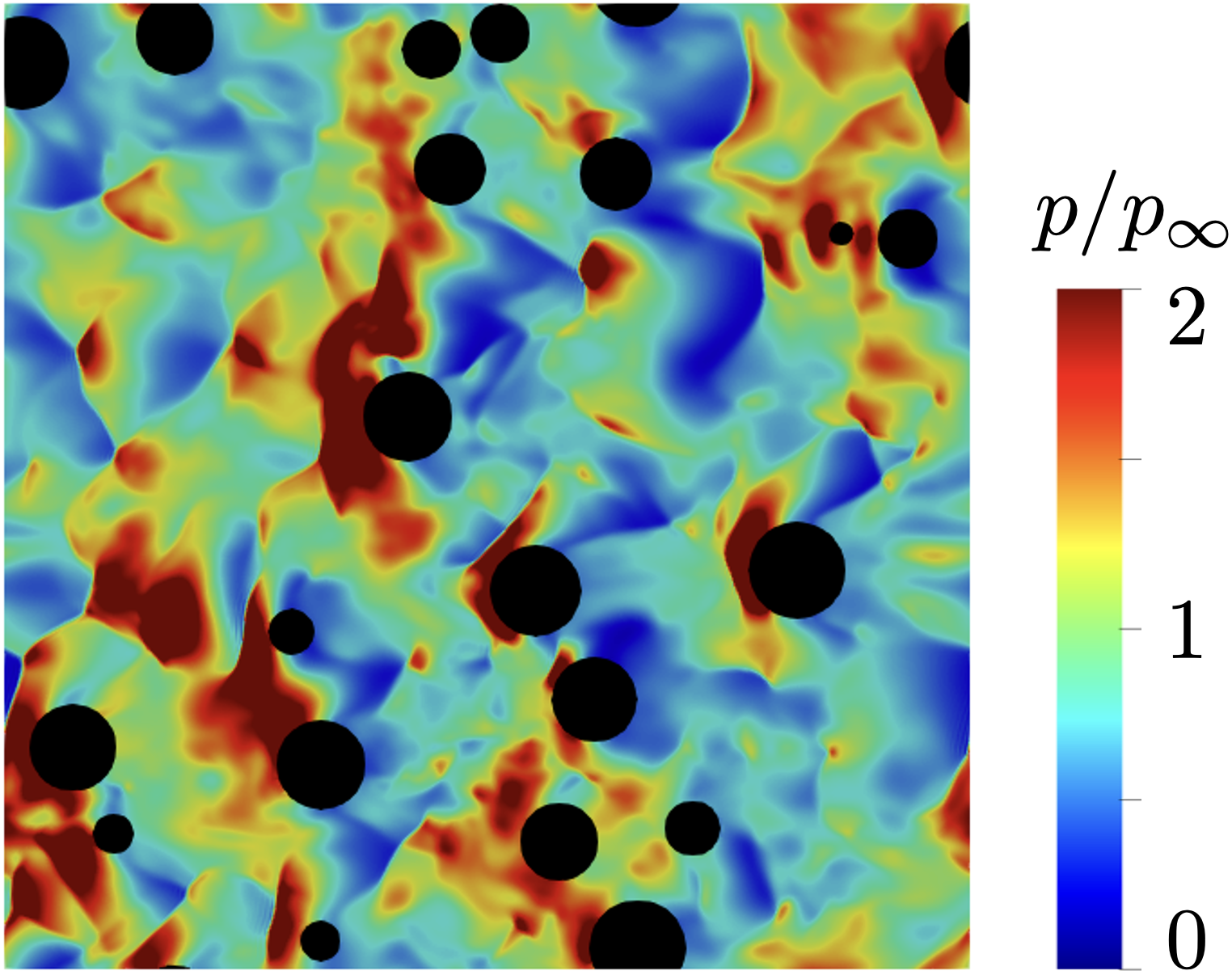}}~~
  \caption{Instantaneous snapshots showing a two-dimensional slice of velocity magnitude (top) and pressure (bottom) with $\phi{=}0.1$ and $M{=}0.1$ (left), $M{=}0.5$ (middle) and $M{=}1$ (right). Particles shown in black.}  
  \label{fig:vel_pres}
\end{figure}
The simulations are run out until a statistical stationary state is reached. For a fixed Reynolds number and volume fraction, qualitative changes in the flow field are observed as the free-stream Mach number varies (see Fig.~\ref{fig:vel_pres}). At low $M$, the velocity field is unsteady with relatively uniform pressure. Wakes shed by particles induce long-range hydrodynamic interactions. As $M$ increases, the nature of these interactions changes--high pressure regions form upstream of particles with the emergence of bow shocks when $M{\approx} 1$. The level of compressibility and mean drag force acting over the particle assembly are quantified in the following sections and a new drag correlation is proposed.

\subsection{Effects of particles on gas-phase compressibility}
The level of compressibility can be quantified by the turbulent Mach number, defined here as $M_t{=}\sqrt{2k}/\overline{c}$ where $k{=}\widetilde{\bm{u}'\bm{\cdot}\bm{u}'}/2$ is the `pseudo' turbulent kinetic energy (PTKE),  $\bm{u}'{=}\bm{u}-\widetilde{\bm{u}}$ and $\widetilde{\bm{u}}{=}(U_\infty,0,0)^{\sf T}$. We adopt the terminology PTKE since velocity fluctuations originate at the particle scale and may be non-zero even in laminar flows, e.g. due to steady wakes~\citep{mehrabadi2015pseudo,shallcross2020volume}. The evolution of PTKE in a statistically homogeneous compressible flow is given by
\begin{equation}\label{eq:PTKE}
    \overline{\rho}\left(1-\phi\right)\frac{{\rm d}k}{{\rm d}t}=\mathcal{P}^p+\mathcal{P}^v+\epsilon^p+\epsilon^v,
\end{equation}
where $\mathcal{P}^p{=}-\langle \bm{u}'\bm{\cdot}(p\bm{n})\delta(\bm{x}-\bm{x}^{(I)})\rangle$ and $\mathcal{P}^v{=}\langle \bm{u}'\bm{\cdot}(\bm{\tau}\bm{\cdot}\bm{n})\delta(\bm{x}-\bm{x}^{(I)})\rangle$ are production terms due to pressure and viscous drag, respectively, $\bm{n}$ is the unit normal vector pointing outward from the surface of the particle, $\delta(\bm{x}-\bm{x}^{(I)})$ is the Dirac delta function that ensures the stress is evaluated at the gas--solid interface, $\epsilon^v {=} (1-\phi)\overline{\bm{\tau} \bm{:} \nabla \bm{u}'}$ represents viscous dissipation and $\epsilon^p {=} -(1-\phi)\overline{p\nabla \bm{\cdot} \bm{u}'}$ is the pressure--strain correlation.

\begin{figure}
    \centering
    \sidesubfloat[]{\includegraphics[height=.22\textheight]{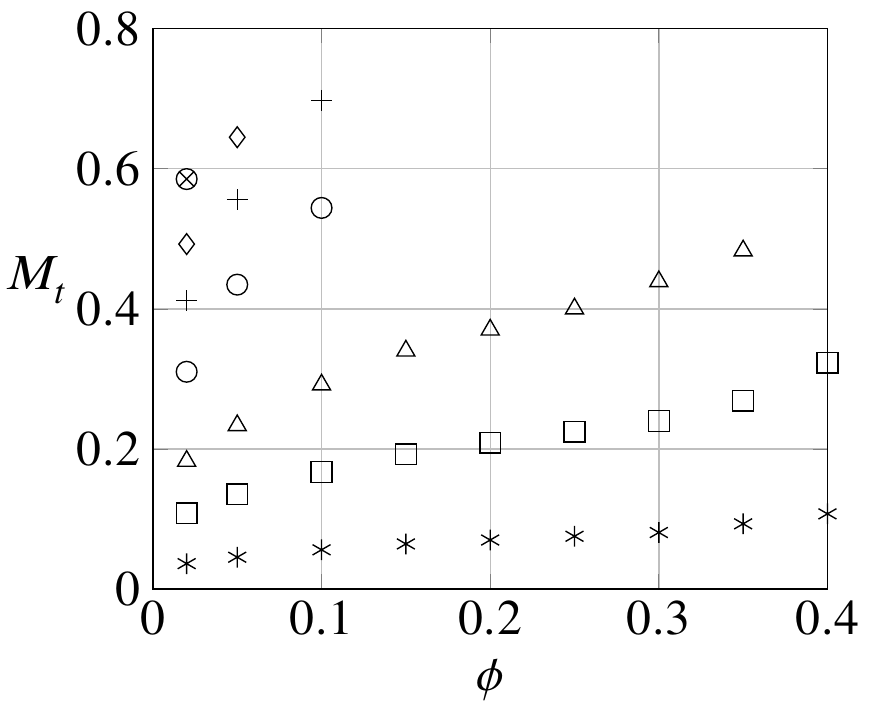}}
    \sidesubfloat[]{\includegraphics[height=.22\textheight]{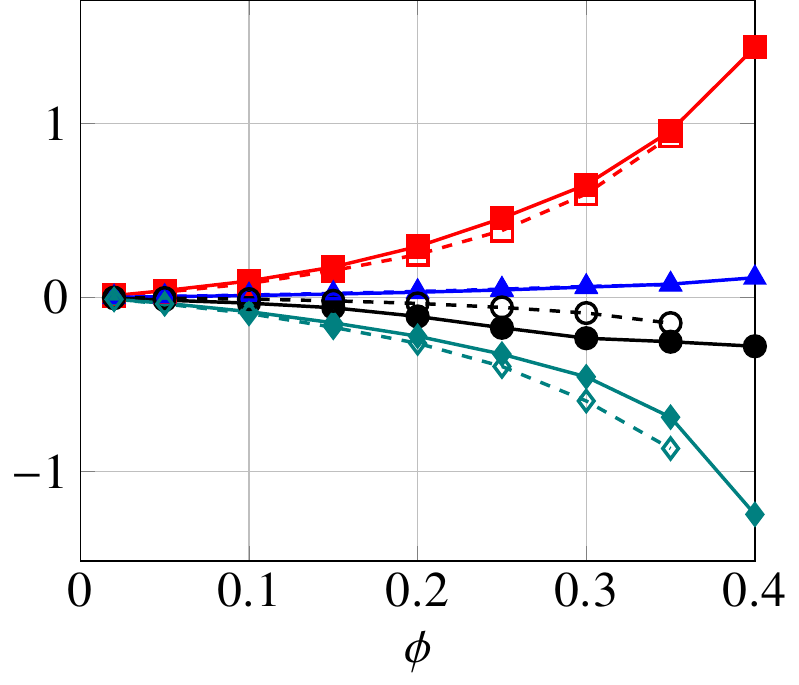}}
    \caption{(a) Turbulent Mach number as a function of volume fraction for $M{=}0.1$ ($\ast$), $M{=}0.3$ ($\square$), $M{=}0.5$ ($\vartriangle$), $M{=}0.8$ ($\bigcirc$), $M{=}1.0$ (+), $M{=}1.2$ ($\lozenge$) and $M{=}1.5$ ($\otimes$). (b) Contributions to the PTKE budget~\eqref{eq:PTKE} normalized by $(1-\phi)\rho_\infty U_\infty^3/D$ with $\mathcal{P}^p$ ($\square$), $\mathcal{P}^v$ ($\triangle$), $\epsilon^p$ ($\bigcirc$), $\epsilon^v$ ($\lozenge$) for $M{=}0.1$ (--; closed symbols) and $M{=}0.5$ (- -; open symbols).}
    \label{fig:turbulentMach}
\end{figure}

From Fig.~\ref{fig:turbulentMach}(a) it can be seen that $M_t$ increases monotonically with increasing $M$ and $\phi$. The PTKE budget (Fig.~\ref{fig:turbulentMach}(b)) shows that drag production, namely through pressure, increases significantly with increasing $\phi$. Thus, the increase in $M_t$ can be attributed to PTKE that originates from neighbour-induced hydrodynamic interactions. Interestingly, the pressure--strain correlation, not present in incompressible flows~\citep{mehrabadi2015pseudo}, plays a role at moderate $\phi$ even at low $M$. The small values of $M_t$ at low $\phi$ and $M$ correspond to flows where the acoustic time scale is much smaller than the convective and viscous time scales, indicative of small dilatational fluctuations. The majority of the simulations presented here are in the nonlinear subsonic regime (moderate $M_t$ with $M{<}1$), characterized by substantial dilatational fluctuations (shocklets). In summary, for a given free-stream Mach number, increasing the particle loading results in larger values of $M_t$ and correspondingly the level of compressibility increases. 

\subsection{Drag statistics}\label{sec:drag}
\begin{figure}
    \centering
      \sidesubfloat[]{\includegraphics[height=.22\textheight]{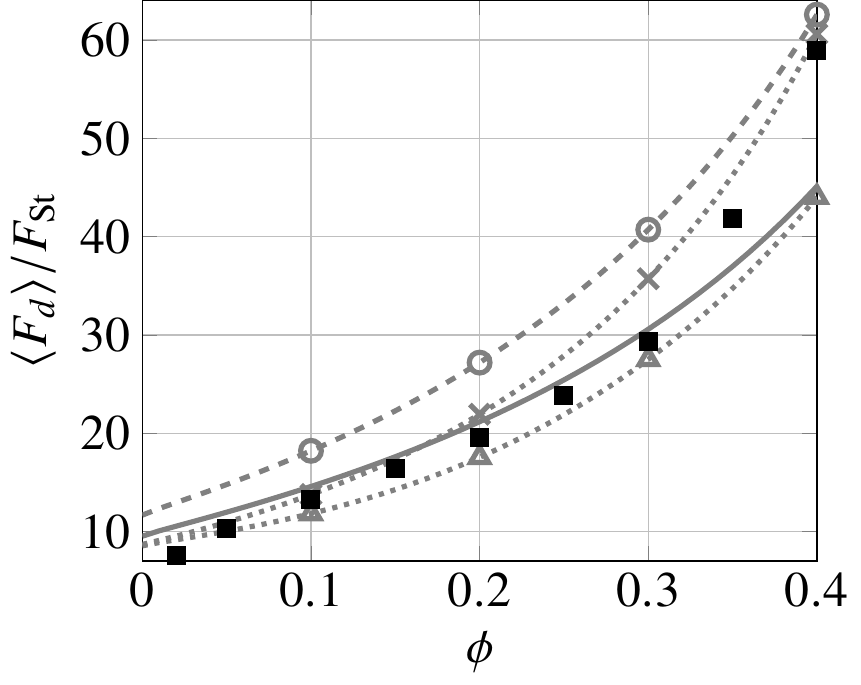}}~
      \sidesubfloat[]{\includegraphics[height=.22\textheight]{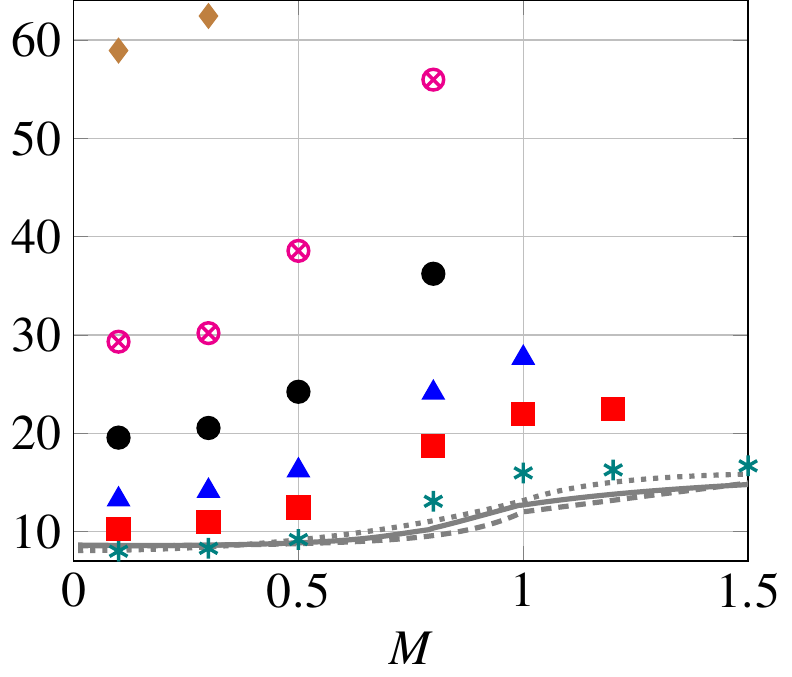}}~
    \caption{Mean drag normalized by Stokes' result as a function of $\phi$ and $M$. (a) $M{=}0.1$ ($\blacksquare$) compared to incompressible correlations of \citet{tenneti2011drag} ($\triangle$), \citet{beetstra2007drag} ($\bigcirc$), \citet{tang2015new} (--) and \citet{ling2012interaction} ($\times$). (b) $\phi{=}0.02$ ($\ast$), $\phi{=}0.05$ ($\square$), $\phi{=}0.1$ ($\vartriangle$), $\phi{=}0.2$ ({\large $\bullet$}), $\phi{=}0.3$ ($\otimes$) and $\phi{=}0.4$ ($\lozenge$) compared to compressible correlations of  \citet{henderson1976drag} (- -), \citet{loth2021supersonic} (--) and \citet{singh2022general} ($\cdots$).}
    \label{fig:drag_compare}
\end{figure}

The drag force averaged over the entire suspension normalized by Stokes' solution ($F_{\rm St}$) is shown in Fig.~\ref{fig:drag_compare}. At low free-stream Mach number ($M{=}0.1$), the results lie within the
range of existing correlations developed for incompressible flows; see Fig.~\ref{fig:drag_compare}(a). 
At low volume fraction the drag force is approximately 10 times greater than Stokes' result due to finite Reynolds number effects. The presence of neighbouring particles results in larger relative drag, with $\langle F_d\rangle/F_{\rm St}{\approx} 60$ when $\phi{=}0.4$.

Figure~\ref{fig:drag_compare}(b) shows how the normalized drag force varies with free-stream Mach number. At low volume fraction ($\phi{\lessapprox}.02$), drag matches closely with existing compressible correlations. In this low volume fraction limit, there is relatively little effect of $M$ on drag until a critical Mach number is reached at $M{\approx}0.8$. Above this value, drag increases sharply with $M$ due to pressure increases across shock waves. In the supersonic range ($M{>}1$), drag becomes almost independent of $M$ again. Drag increases monotonically with increasing $\phi$ and the critical Mach number decreases due to added compressibility effects caused by particles. Like the isolated particle case, drag becomes nearly independent of $M$ when $M$ is large, though more data is needed to assess trends in drag at combinations of large $M$ and $\phi$.

\subsection{New drag correlation}
Based on the results presented above, we propose a correlation that captures the complex dependency of $\Rey$, $M$ and $\phi$ on the drag coefficient, expressed compactly as
\begin{equation}\label{eq:CD}
    C_D(\Rey,M,\phi)=C_{D,\text{iso}}(Re,M^*)f(\phi).
\end{equation}
The drag correlation for an isolated particle is taken from the continuum regime formulation of \citet{singh2022general}, given by 
\begin{equation}\label{eq:CDiso}
C_{D,\text{iso}}(\Rey,M) =
\left\{
\begin{array}{ll}
C_0\left[1+\delta_0/\sqrt{\Rey}\right]^2 &\textrm{if } M=0  \\
C_0\Theta(M)\left[1+\delta_0/\sqrt{\widehat{\Rey}}\right]^2 & \textrm{if } 0<M\le1 \\
C_0\Theta(M_s)\left[1+\delta_0/\sqrt{\widehat{\Rey}_s}\right]^2+C_1\left[1-a U_s/U_\infty\right] & \textrm{if } M>1. \\
\end{array}
\right.
\end{equation}
We note that other correlations could be used~\citep[e.g.][]{henderson1976drag,parmar2010improved,loth2021supersonic}, though this expression was found to have the least reliance on empirical coefficients. In the subsonic regime, $C_0{=}24/\delta_0^2$ and $\delta_0{=}9.4$~\citep{abraham1970functional} so that $C_{D,\text{iso}}$ reduces to Stokes' result at low $\Rey$ when $M{\rightarrow}0$. A boundary layer transformation from an incompressible to weakly compressible flow is applied via $\Theta(M){=}\left(1+.5(\gamma-1)M^2\right)^{\gamma/(\gamma-1)}$ and $\widehat{\Rey}{=}\Rey\Theta(M)^{(\gamma+1)/2\gamma-n(\gamma-1)/\gamma}$, where $n{=}0.74$ is the viscosity power-law exponent. The subscript $s$ in the supersonic regime denotes post-shock conditions obtained using the Rankine--Hugoniot relations. The coefficient $C_1$ captures increased drag due to pressure changes across the shock, given by $C_1{=}(C_D^{\infty} - C_0 (1+(\gamma-{1})^2/4\gamma)^{\gamma/(\gamma{-}1)}/(1-(\gamma-1)/(a_0 M(\gamma{+}1))$, where $a{=}(1+a_0 (M -1))^{-1}$ and $a_0{=}0.3555$. When $M{\gg}1$, the drag coefficient tends to $C_D^{\infty}{=}0.9$. A detailed derivation of $C_{D,\text{iso}}$ and extension to rarefied flows can be found in \citet{singh2022general}.

\begin{figure}
    \centering
    \sidesubfloat[]{\includegraphics[height=.22\textheight]{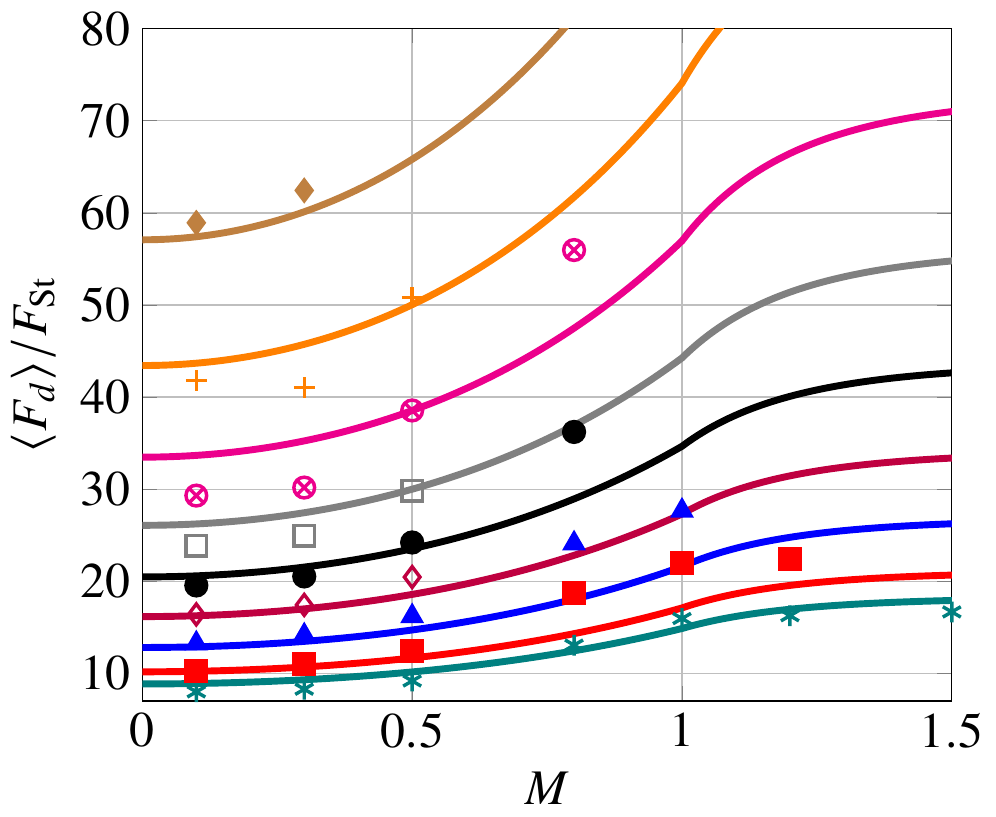}}~
    \sidesubfloat[]{\includegraphics[height=.22\textheight]{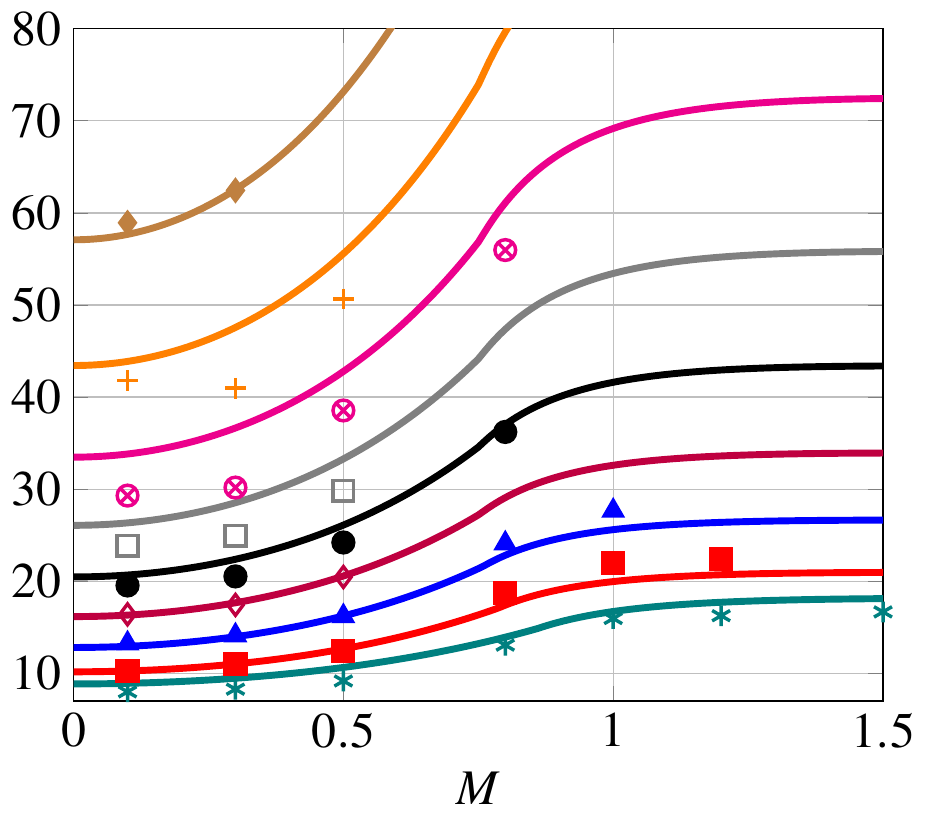}}~
    \caption{Mean drag normalized by Stokes' result as a function of $M$ for different $\phi$. Simulation results (symbols) and correlation given by Eq.~\eqref{eq:CD} (lines) without (a) and with (b) the effective Mach number. $\phi{=}0.02$ ($\ast$; green), $\phi{=}0.05$ ($\square$; red), $\phi{=}0.1$ ($\vartriangle$; blue), $\phi{=}0.15$ ($\lozenge$; brown), $\phi{=}0.2$ ({\large $\bullet$}; black), $\phi{=}0.25$ ($\square$; gray), $\phi{=}0.3$ ($\otimes$; pink), $\phi{=}0.35$ (+; orange), $\phi{=}0.4$ ($\lozenge$; brown).}
    \label{fig:drag_vs_mach}
\end{figure}

To capture the increase in drag due to the presence of neighbouring particles, $C_D$ is augmented with a volume fraction correction $f(\phi){=}\left(1{+}2\phi\right)/\left(1{-}\phi\right)^2$ that was analytically derived from periodic arrays of spheres in the Stokes regime~\citep{sangani1991added}. The observed reduction in critical Mach number with increasing $\phi$ is accounted for by replacing $M$ in $C_{D,\text{iso}}$ with a volume fraction-dependent \emph{effective} Mach number, $M^*$. Simulations reveal a collapse in $M^*$ when normalized by the free-stream Mach number according to
\begin{equation}
    \frac{M^*}{M}=1+\alpha\left(1-e^{-\beta\phi} \right),
\end{equation}
where $\alpha{=}0.33$ and $\beta{=}34$. 
As shown in Fig.~\ref{fig:drag_vs_mach}, the normalized drag force, $\langle F_d\rangle/F_{\rm St}{=}C_D\Rey/24$, is well predicted by the correlation given by Eq.~\eqref{eq:CD} over the range of $M$ and $\phi$ considered. Replacing $M$ with $M^*$ in $C_{D,\text{iso}}$ is needed to capture the correct shift in critical Mach number, especially at intermediate $\phi$.

\section{Conclusions}
The drag force of compressible homogeneous flows past random arrays of spheres with adiabatic boundary conditions was studied using particle-resolved simulations. All of the simulations performed in this work considered a single value of Reynolds number at $\Rey{\approx}300$. The simulations span subsonic to supersonic flow at low to moderate volume fractions and subsonic flow at moderate to high volume fractions. It should be noted that the quasi-steady drag is unlikely to experience combinations of high $\phi$ \textit{and} high $M$ due to the amount of power required to sustain such high-speed flow through a dense suspension.

Neighbour-induced hydrodynamic interactions generate pseudo-turbulent kinetic energy via pressure drag, giving rise to moderate turbulent Mach numbers, even when $M{<}0.3$.
Increasing the volume fraction is found to both increase the magnitude of the drag force and reduce the critical Mach number that demarcates the transonic regime. Despite the high dimensional space, we identified a relatively simple correlation that captures the complex interplay between $M$ and $\phi$ on drag force. In the limit $\phi{\ll}1$, the correlation reduces to the continuum formulation of \citet{singh2022general} that is valid for $Re{<}10^4$ and $M{<}6$. The Mach number used in the correlation is replaced by a volume fraction-dependent effective Mach number to capture the shift in critical Mach number. Future work should focus on different thermal boundary conditions and the role of Reynolds number on  the effective Mach number and the volume fraction correction.

\backsection[Acknowledgements]{Resources supporting this work were provided by the NASA High-End Computing (HEC) Program through the NASA Advanced Supercomputing (NAS) Division at Ames Research Center.}
\backsection[Funding]{This work was supported by the National Aeronautics and Space Administration (NASA) grant no. 80NSSC20K0295.}
\backsection[Declaration of interests]{The authors report no conflict of interest.}
%
%
\backsection[Author ORCID]{M. Khalloufi, https://orcid.org/0000-0002-5790-3484; J. Capecelatro, https://orcid.org/0000-0002-1684-4967}
%
%
%
%
%
%
%
%
\bibliographystyle{jfm}
\bibliography{references}

\end{document}